\documentclass[aps]{revtex4}
\usepackage{times}
\usepackage{epsfig}
\usepackage{amssymb}
\usepackage{bm}
\usepackage{amsmath}
\usepackage{amsfonts}

\begin{document}

\title{Local burst model of CMB temperature fluctuations: luminescence in lines of primary para- and orthohelium}

\author{V. Dubrovich$^{1}$, S. Grachev$^{2}$, T. Zalialiutdinov$^{3}$}
\affiliation{ 
$^{1}$ Special Astrophysical Observatory, St. Petersburg Branch, Russian Academy of Sciences, 196140, St. Petersburg, Russia \\
$^{2}$ Sobolev Astronomical Institute, St. Petersburg State University, Universitetskii pr. 28, St. Petersburg, 198504 Russia\\
$^{3}$ Department of Physics, St. Petersburg State University, Petrodvorets, Ulianovskaya 1, 198504, St. Petersburg, Russia}

\begin{abstract}
We have considered the formation of the luminescent subordinate HeI lines by the absorption of continuum radiation from a source in the lines of the main HeI series in the expanding Universe. It is suggested that at some moment of time, corresponding to the redshift $z_{0}$, a burst of superequilibrium blackbody radiation with a temperature $ T+\Delta T $ occurs. This radiation is partially absorbed at different $z <z_{0}$ in the lines of the main HeI series and then converted into the radiation of subordinate lines. If $ \nu_{ij} $ is the laboratory frequency of the transition of some subordinate line originating at some $z$, then in the present time its frequency will be $ \nu=\nu_{ij}/(1+z) $. For different $z$ (and, consequently, for different $  \nu $), the quantum yield for the subordinate lines of para- and orthohelium - the number of photons emitted in the subordinate line, per one initial excited atom and line profiles are calculated. Different pumping channels were considered. Spatial and angular distributions of radiation intensity of luminescent lines for the spherically symmetric radiation sources are presented. It is shown that for sufficiently large $ \Delta T/T $, the luminescent lines can be very noticeable in the spectrum of blackbody background radiation.
\end{abstract}

\maketitle

\section{Introduction}
\label{intro}
In recent decades, the theoretical study of the epoch of primary recombination in the early Universe has been of particular interest in connection with high-precision observations of the temperature and polarization anisotropy of the cosmic microwave background (CMB). To date, an extensive theoretical base has been developed and a large amount of experimental material has been achieved in this direction \cite{planck2015}. A detailed study of the processes leading to the spectral-spatial distortions of the CMB is of particular interest \cite{chluba, kurt}. In particular, medium-resolution spectroscopy of individual objects on the CMB brightness map seems very important. What is new here is the transition from investigating the statistical CMB characteristics to searching for and studying local phenomena and "protoobjects". The latter can be quite rare events that have virtually no influence on the average statistical CMB parameters, but at the same time they give information about completely new physical laws.  Apart from more or less probable but still hypothetical objects, there is definitely a whole class of local sources in the early Universe that can be individually investigated. These are the same standard primordial CMB temperature fluctuations that are being studied with such statistical care today. In fact, we are dealing with some regions of space where there is a temperature increase or decrease for some reason. It is very important that apart from the spatial isolation of these regions, the temperature deviation in them is, in addition, nonstationary. Depending on the formation mechanism of this inhomogeneity, the characteristic time of its growth and decay can be different. With regard to such events we can talk about a local burst model of fluctuations. 

The main physical assumptions used in this paper are as follows. In the Universe at all stages of its evolution there are fluctuations of matter density and radiation. In our case, in the standard theory of evolution, these fluctuations arise as a result of the chaotic addition of weak sound waves in a radiation-dominated hydrogen-helium plasma. The speed of these waves is about half the speed of light. Therefore, these fluctuations in a given region of space can be considered as fast-changing sources by analogy with galaxies, quasars immersed in the intergalactic medium (hydrogen-helium plasma). These "sources" have a size determined by the spectrum of the scales of sound waves, the lifetime, intensity and the radiation spectrum. The lifetime of such sources is very small because of the high speed of sound. Physically, this picture is similar to the picture of boiling water. This radiation (the super-equilibrium part) under propagation in the surrounding homogeneous medium from the source to the observer undergoes scattering on free electrons, in lines of atoms and ions, and redistribution of energy over the spectrum due to the luminescence mechanism.

It is important to optimize the experimental search of new effects using the magnitude and range of wavelengths. In connection with this, it becomes important to consider second-order effects in the inelastic scattering of superequilibrium photons. The effect we have considered here is essential for the search of primary CMB distortions formed at large redshifts. In previous works \cite{dubr2016,2016} the effects of inelastic scattering in the resonance lines of hydrogen and helium were described. The resulting distortions of the CMB spectrum lie in the distant wing. This makes the task of detecting and investigating them in detail very difficult and almost impossible to observe with ground-based equipment. In this work we present a calculation of quantum yield in resonance lines for the luminescence mechanism of excitation, which leads to the formation of CMB distortions at the frequencies of subordinate transitions. These frequencies can be almost two or three orders of magnitude smaller than the original resonant frequencies, and the corresponding CMB distortions are in the centimeter and millimeter wavelength ranges at the present time. Ground observations in these ranges are available and do not require extraordinary efforts, which significantly increases the probability of their detection. 

In order to calculate profiles of HeI luminescent lines arising due to absorption of continuum radiation from a source in the lines of main series one has to solve the problem of radiation transfer in these lines. In the range of redshifts $z=3000 - 4500$, where the luminescence of HeI lines occurs, an optical thickness of the Universe in the first line of the main series (transition $ 1^1S-2^1P $) changes from $10^4$ to $1$. On the other hand the Universe is practically transparent for the lines of subordinate series. So, the problem is formulated as follows. At some moment of time corresponding to redshift $z_0$ a source of continuum radiation arises (bursts). This radiation begins to propagate in space undergoing scatterings on atoms and electrons. A part of radiation that is absorbed by atoms in the lines of the main series converts into subordinate lines radiation at each $z\leqslant z_0$. If we neglect the width of absorption coefficient profile regarding it as a delta function $\delta(\nu-\nu_{ij})$ in comoving frame of reference, where $\nu_{ij}$ is a frequency of transition $i\leftrightarrow j$, then $q_{ij}$ subordinate photons with frequency $\nu_{ij}$ arises per one photon absorbed in the line of main series. Furthermore, these subordinate photons undergo only Thomson scattering on electrons (without frequency change), spatial dilution and red shifting due to the Universe expansion. Finally they are registered at the present day epoch ($z=0$) on the frequency $\nu=\nu_{ij}/(1+z)$. As a result an observed profile of a subordinate line forms. Each point of the profile corresponds to a certain redshift and its width is much greater then the width of absorption coefficient profile.

The method used in the present work for the estimations of quantum yield $ q_{ij} $ was first proposed in \cite{bernberndubr} for an arbitrary type of initial CMB distortions and applied later in \cite{dubrlipka, dubr1997} for the calculations of CMB distortions from primary molecules. By definition, the quantum yield $q_{ij}$ is the ratio of mean number of photons of specified frequency emitted in transition from the upper level $i$ to the lower level $j$ to the number of resonant photons in the pumping channel. Then the luminescence intensity is proportional to the pumping intensity and to the quantum yield $q_{ij}$ depending on the nature of the trajectory of the excited electron over the levels. Below we extend the approach proposed in \cite{2016} for the calculations of luminescence intensity of para- and orthohelium subordinate lines with the account of spin-flip transitions between triplet and singlet helium states. 

The paper is organized as follows. Section \ref{method} is devoted to the calculation of quantum yield $ q_{ij} $ in the subordinate lines of para- and orthohelium. In Section \ref{method2} the problem of radiation transfer of HeI lines is considered. Spatial and angular distributions of radiation intensity of luminescent lines for the spherically symmetrical radiation sources are presented. Results of numerical calculations and discussion of them are given in Section \ref{res}. 

\section{Quantum yield in the subordinate lines}
\label{method}

The following model of the formation of luminescent lines is considered. At a certain moment in time, corresponding to the redshift $z_0$, an instantaneous burst of isotropic radiation in the continuum occurs in the expanding Universe \cite{dubr2015,dubr2016}. This radiation is absorbed at different redshifts $z$, giving rise to luminescent subordinate lines. 

Let $ \nu_{ij} $ be the laboratory frequency of the transition arising at some $ z $ in certain subordinate line $ i\rightarrow j $. Then taking into account the redshift, the frequency of this transition is $\nu = \nu_{ij} / (1 + z)$. We assume that $z$ is large enough that the medium is in local thermodynamical equilibrium (LTE), and the population of the levels is determined by the Boltzmann and Saha equations. Additional population of the $ k $ level due to the absorption of radiation from the source leads to a nonLTE distribution for the population of atomic level, that is, spreading to other levels under the action of background radiation. As a result of such a process, a certain amount of super-equilibrium radiation in subordinate lines escapes from the matter. In other words, a luminescence occurs.

The intensity of the luminescent lines is proportional to the quantum yield. The quantum yield can be found from the solution of the system of kinetic balance rate equations
\begin{eqnarray}
\label{rate1}
\frac{dn_{i}}{dt}=-n_{i}\sum_{j=1}^{N}R_{ij}+\sum_{j=1}^{N}n_{j}R_{ji}
\end{eqnarray}
where $ n_{i} $ is the occupation number of $ i $-th level, $ R_{ij} $ is the probability coefficient (transition rate) for the transition $ i\rightarrow j  $, $ t $ is the time, $ N $ is the number of considered bound states. Sum over $ i $ in Eq. (\ref{rate1}) implies summation over the set of quantum numbers $ n_{i}S_{i}L_{i} $, where $ n_{i} $ is the principal quantum number, $ S_{i} $ is the total spin and $ L_{i} $ is the angular momentum. Transition rates in the presence of blackbody background radiation are given by the following relations
\begin{eqnarray}
\label{2}
R_{ij}=\frac{A_{ij}}{1-\mathrm{exp}\left[-\frac{h\nu_{ij}}{k_BT}\right]}
,\end{eqnarray}
for the transitions from the upper level to the lower one ($ E_{i}>E_{j} $) and 
\begin{eqnarray}
\label{3}
R_{ij}=\frac{g_{j}}{g_{i}}\frac{A_{ji}}{\mathrm{exp}\left[\frac{h\nu_{ij}}{k_BT}\right]-1}
\end{eqnarray}
for transitions from the lower level to the upper one ($ E_{i}<E_{j} $). Here $ \nu_{ij}=\left|E_i-E_j \right| $ is transition frequency in the laboratory frame of reference, $ A_{ij} $ is the Einstein coefficient for the probability of spontaneous emission, $ g_{i} $ is the statistical weight of state $ i $, $ k_B $ is the Boltzmann constant and $ T $ is the radiation temperature associated with the redshift parameter $ z $ by the relation $ T=T_{0}(1+z) $, where $ T_{0} = 2.725$ K is the present CMB temperature. 

It is convenient to rewrite Eq. (\ref{rate1}) in terms of Menzel factors \cite{kaplan}
\begin{eqnarray}
\label{menzel}
b_{i}=n_{i}/n^{\rm LTE}_{i},
\end{eqnarray}
where $ n^{\rm LTE}_{i} $ is the equilibrium population of level $ i $ given by Saha equation
\begin{eqnarray}
\label{saha}
n^{\rm LTE}_{i}=n_{e}n_{c}\frac{g_{i}}{4}
\left(\frac{2\pi m_{e}k_{B}T}{h^2}\right)^{-3/2}e^{E_{i}^{\rm ion}/k_{B}T}.
\end{eqnarray}
Here $ n_{e} $ is the electron number density, $ n_{c} $ is number density of helium nucleus, $ h $ is the Planck constant, $ m_{e} $ is the electron mass and $ E_{i}^{\rm ion} $ is the ionization energy for electron in state $ i $. Then using Eq. (\ref{menzel}) and taking into account that $ n_{i}^{\rm LTE}R_{ij}=n_{j}^{\rm LTE}R_{ji} $ system of rate Equations (\ref{rate1}) takes the form
\begin{eqnarray}
\label{rate2}
\frac{db_{i}}{dt}=-\sum_{j=1}^{N}R_{ij}(b_{i}-b_{j})
.\end{eqnarray}
We looked for corrections to equilibrium populations, meaning that we sought the solution of Equations (\ref{rate2}) in the form $1 + \Delta b_{i}$. Obviously, the system of equations for corrections $ \Delta b_{i} $ has the same form as for the populations themselves. Therefore below we will understand $ b_{i} $ as corrections to populations. For the boundary conditions we should set the Menzel factors at initial moment of time $ t=0 $. Since luminescence arises upon absorption of radiation in the line $ 1^1S\rightarrow k^{2S_{k}+1}P $ ($ k=2,\,3,\,4\dots $) these conditions are: $ b_{i}(0)=0 $ if $ i\neq k $ and $b_{k}(0)=1$ if $ i=k $. Then the system of rate Equations (\ref{rate2}) can be solved numerically. We note that in Eq. (\ref{rate2}) we neglect bound-free transitions. This approximation can be justified by the different initial conditions for the problem of luminescence, in contrast to the problem of ordinary recombination \cite{bernberndubr}. For the latter case these conditions are: $ b_{i}(0)=\alpha_{i} $ ($ \alpha_{i} $ is the recombination coefficient to the state $ i $). In the case of luminescence only a small part of atoms excited at $ t=0 $ to state $ k^{2S_{k}+1}P  $ ($ k=2,\,3,\,4\dots $) can be again ionized and therefore bound-free transitions play a less important role for our estimations of quantum yield in subordinate lines.

By the definition, the quantum yield in the transition between the upper level $ i $ and the lower level $ j $ is the number of uncompensated transitions in this line per one initial excited atom in the pumping line. Since we assumed that the Menzel factor for the upper level of pumping line is $ b_{k}=1 $ at $ t=0 $, then the population of this level is $n_{k}^{\rm LTE}$, i.e., given by Saha Equation (\ref{saha}). Finally the number of uncompensated transitions (quantum yield) in line $ i\rightarrow j $ is obtained by multiplying corresponding term in system (\ref{rate2}) by $ n_{i}^{\rm LTE} $
\begin{eqnarray}
\label{escape}
q_{ij}=\frac{n_{i}^{\rm LTE}}{n_{k}^{\rm LTE}}R_{ij}\int\limits^{\infty}_{0}\left(b_{i}(t)-b_{j}(t) \right)dt
.\end{eqnarray}
During the establishment of equilibrium after an initial population, transitions to the ground state in the lines of the main series can occur. The optical thickness of the medium can be very large for these transitions in contrast to the subordinate lines. A photon in the line of the main series can scatter many times before either "dying" due to fragmentation into photons in subordinate lines, or leaving the scattering process due to Doppler shift in the line wing as a result of the Universe expansion. The latter circumstance can be taken into account by multiplying the corresponding term in Eq. (\ref{rate2}) by the Sobolev escape probability \cite{seager2000}
\begin{eqnarray}
\label{sobolev}
p_{ji}=\frac{1-e^{-\tau}}{\tau},
\end{eqnarray}
where $\tau$ is the optical length in the line
\begin{eqnarray}
\label{tau}
\tau=\frac{A^{\rm E1}_{ij}\lambda_{ij}^3}{8\pi H(z) }\left[n_{j}\frac{g_{i}}{g_{j}}-n_{i}\right].
\end{eqnarray}
Here $\lambda_{ij}$ is the transition wavelength, $c$ is the speed of light, $H(z)$ is the Hubble factor:
\begin{eqnarray}
\label{hubble}
H(z)=H_{0}\left(\Omega_{\Lambda}+\Omega_{\mathrm{K}}(1+z)^2
+
\Omega_{\mathrm{m}}(1+z)^3
\right.
\\\nonumber
\left.
+\Omega_{\mathrm{rel}}(1+z)^4\right)^{1/2}.
\end{eqnarray}  
Parameters $H_{0}$, $\Omega_{\Lambda}$, $\Omega_{\mathrm{m}}$, $\Omega_{\mathrm{rel}}$ in Eq. (\ref{hubble}) are listed in Table~\ref{param} and, following \cite{seager2000}, we consider $\Omega_{\mathrm{tot}}= \Omega_{\Lambda}+\Omega_{\mathrm{K}}+\Omega_{\mathrm{m}}+ \Omega_{\mathrm{rel}}=1$ and $\Omega_{\mathrm{K}}=0$. For the ratio of populations in square brackets in Eq. (\ref{tau}), the Boltzmann equation was used. The population of the ground state was calculated from the Boltzmann and Saha formulae. Calculations showed that for large $ \tau $, accounting for opacity in the lines of the main series significantly affects the quantum yield in the subordinate lines. 

We considered a model of a helium atom with $15$ singlet states and $ 14$ triplet states ($ N=29 $ in Eq. (\ref{rate2})) with main quantum numbers $ n\leqslant 5$ and angular momenta $ L\leqslant 4 $. The spin-flip electric dipole transitions (E1) in subordinate lines were taken into account as well as two-photon $ 2^1S\rightarrow 1^1S+2\gamma(\mathrm{E1}) $ and one-photon $ 2^3P\rightarrow 1^1S+\gamma(\mathrm{E1}) $ transitions. The wavelengths, Einstein coefficients and ionization energies were taken from \cite{weise2009,drakebook,timur}. 
All results were checked for convergence. 

\begin{figure}[ht]
\caption{Quantum yield $ q $ in the line $ 2^1P-2^1S $ for the different pumping channels $ 1^1S\rightarrow k^1P $  ($ k=2,3,4,5 $) with the accounting for spin-flip transitions. The lower horizontal axis is a redshift, the upper one is the frequency $ \nu=\nu_{2^1P,2^1S}/(1+z) $ in GHz, where $ \nu_{2^1P,2^1S} $ is the laboratory transition frequency.}
\centering
\includegraphics[scale=0.62]{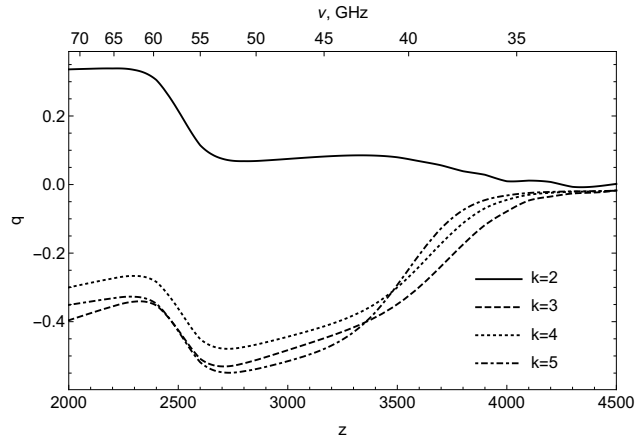}
\label{fig1}
\end{figure}

\begin{figure}[ht]
\caption{Relative difference $ \Delta q/q=(q-q')/q' $ for the quantum yield in the line $ 2^1P-2^1S $ calculated with ($ q $) and without ($ q' $) spin-flip transitions. Four different pumping channels $ 1^1S\rightarrow k^1P $ ($ k=2,3,4,5 $) are presented.}
\centering
\includegraphics[scale=0.62]{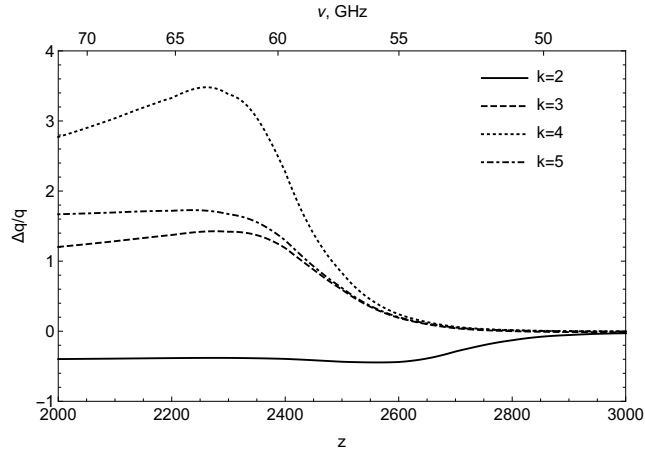}
\label{relative}
\end{figure}

We find that the accounting for spin-flip transitions, particularly $ 2^3P\rightarrow 1^1S+\gamma(\mathrm{E1}) $ and $ 2^3P\rightarrow 2^1S+\gamma(\mathrm{E1}) $, significantly affects the quantum yield in $ 2^1P-2^1S $ line within the interval $ z=2000-2700 $, see Fig. \ref{fig1}. The relative difference for the quantum yield in this line calculated with and without the accounting for spin-flip transitions is presented in Fig. \ref{relative}. We can conclude that efficiency of absorption in $ 2^1P-2^1S $ line with initial pumping in $ 3^1P $,  $ 4^1P $ or $ 5^1P $ states rise up for $ z=2000-2700 $ in comparison with previous calculations \cite{2016}. It is important to note that this period corresponds to redshifts where recombination HeI occurs.

\begin{figure}[ht]
\caption{Quantum yield $ q $ in the line $ 2^3P-2^3S $ for pumping channel $ 1^1S\rightarrow 2^3P $ with the accounting for spin-flip transitions. The lower horizontal axis is a redshift, the upper one is the frequency $ \nu=\nu_{2^3P,2^3S}/(1+z) $ in GHz, where $ \nu_{2^3P,2^3S} $ is the laboratory transition frequency.}
\centering
\includegraphics[scale=0.62]{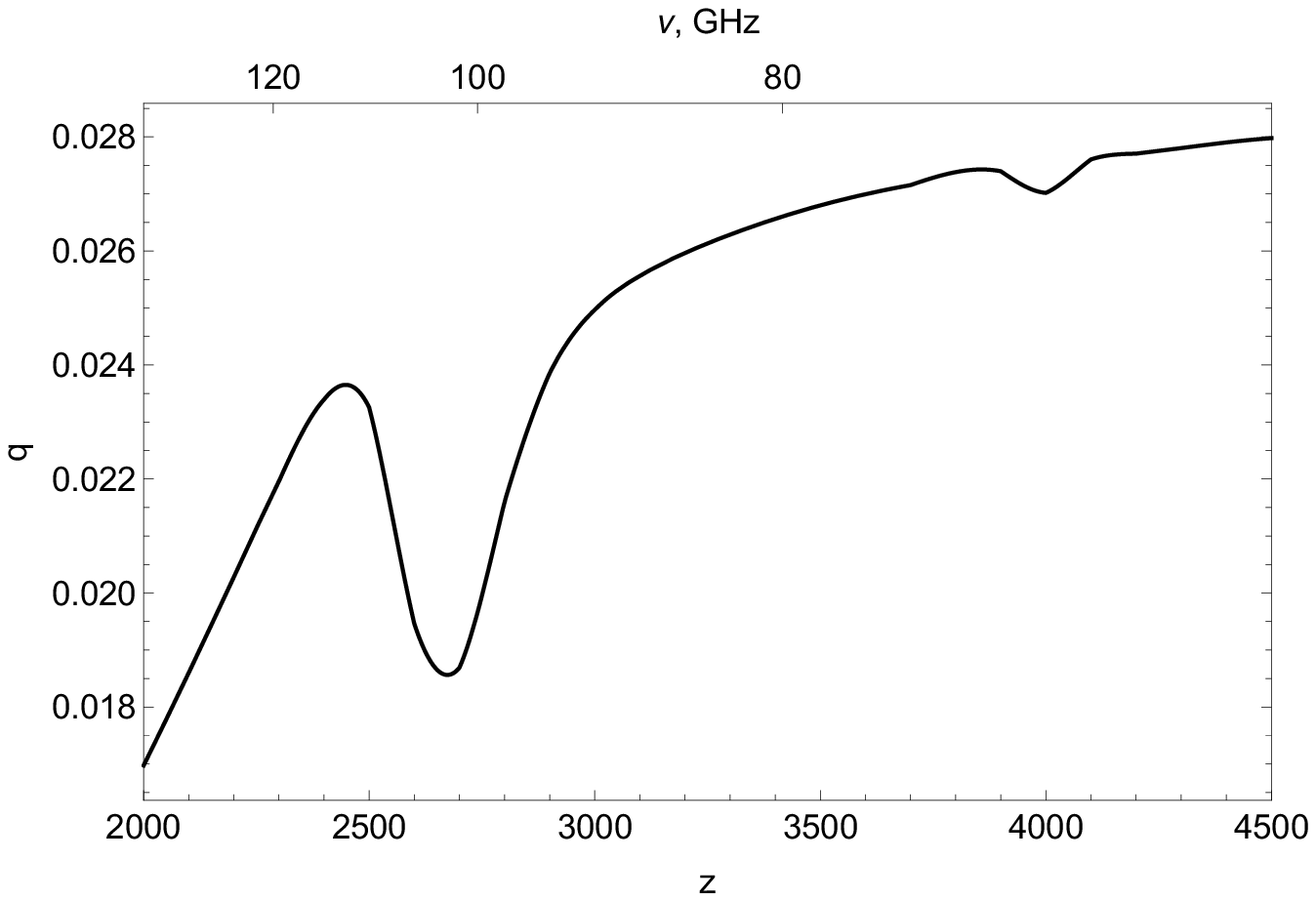}
\label{fig3}
\end{figure}

\begin{figure}[ht]
\caption{Quantum yield $ q $ in the line $ 2^3P-2^3S $ for the different pumping channels $ 1^1S\rightarrow k^3P $  ($ k=3,4,5 $) with the accounting for spin-flip transitions. The lower horizontal axis is a redshift, the upper one is the frequency $ \nu=\nu_{2^3P,2^3S}/(1+z) $ in GHz, where $ \nu_{2^3P,2^3S} $ is the laboratory transition frequency.}
\centering
\includegraphics[scale=0.62]{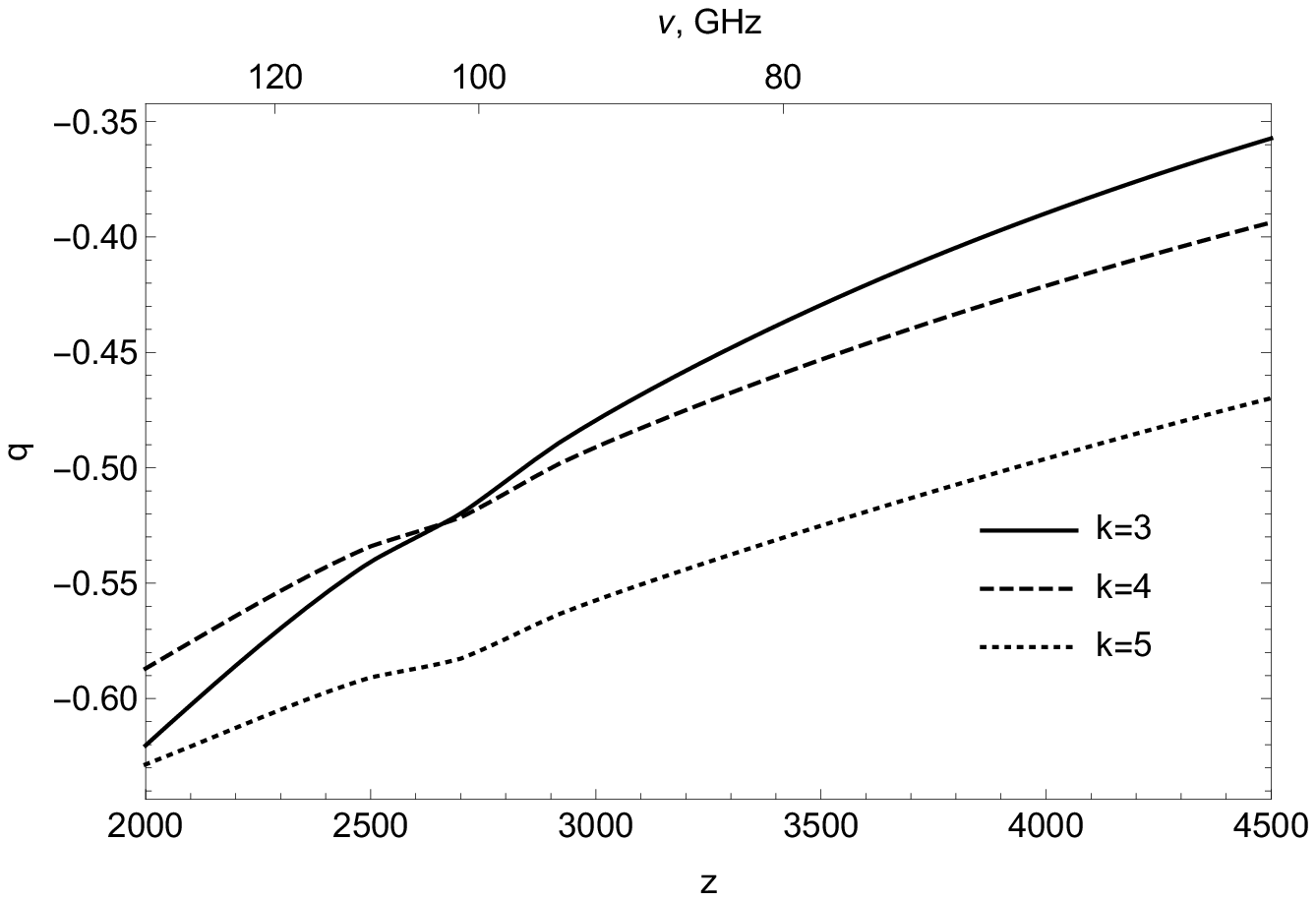}
\label{fig_add}
\end{figure}

\begin{figure}[ht]
\caption{Quantum yield $ q $ in the line $ 4^1D-3^1P $ for pumping channel $ 1^1S\rightarrow 3^1P $ with the accounting for spin-flip transitions. The lower horizontal axis is a redshift, the upper one is the frequency $ \nu=\nu_{4^1D,3^1P}/(1+z) $ in GHz, where $ \nu_{4^1D,3^1P} $ is the laboratory transition frequency.}
\centering
\includegraphics[scale=0.62]{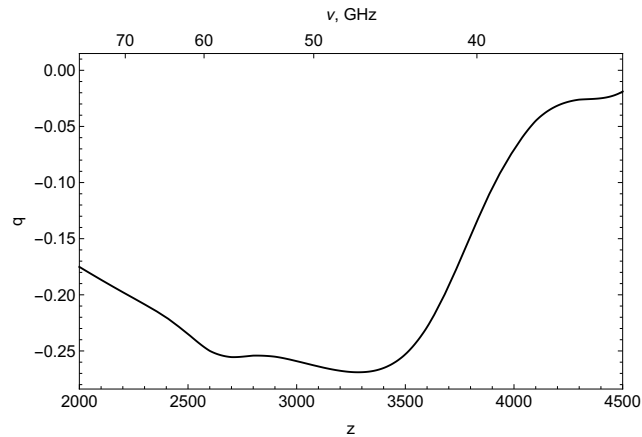}
\label{fig4}
\end{figure}

In Figs. \ref{fig3}-\ref{fig4} we present the quantum yield in different subordinate lines which is the same order as in $ 2^1P-2^1S $ line (Fig. \ref{fig1}). The quantum yield in other HeI subordinate lines is found to be several orders of magnitude smaller and therefore is not of our interest.

\begin{table}
\caption{Parameters of the cosmological model used in the present calculations \cite{planck2015}.}
\label{param}
\begin{tabular}{ c  c  c  }
\hline
\hline
Description & Symbol & Value \\
\hline
Total density & $\Omega_{\mathrm{tot}}$ & 1 \\
Dark energy density & $\Omega_{\Lambda}$ & 0.6911 \\
Baryon density & $\Omega_{\mathrm{b}}$  & 0.0486 \\
Dark matter density & $ \Omega_{\mathrm{c}} $ & 0.2589 \\
Nonrelativistic matter density & $\Omega_{\mathrm{m}}$  & 0.3089 \\
Relativistic matter density & $\Omega_{\mathrm{rel}}$  & $9.07\times 10^{-5}$\\
Hubble constant & $ H_{0} $ & 67.74 $\frac{\rm km}{\rm s\;Mpc}$\\
Present temperature of CMB & $T_{0}$ & 2.725  K\\
Hydrogen mass fraction & X & 0.76\\ 
Helium mass fraction & Y & 0.24\\ 
\hline    
\end{tabular}
\end{table}

\section{Radiative transfer in HeI lines}
\label{method2}

In the considered model of local bursts an influence of spatial dilution and multiple Thomson scattering should be taken into account twice: firstly for propagation of a source continuum radiation from the moment of burst (at $z_0$) to the moment of absorption (at some $z<z_0$) in the lines of the main series and secondly for propagation of subordinate line radiation (arising at $z$ as a result of luminescence) from the moment $z$ to the moment $z=0$. But as we see, an interval of $z$ defining the main part of the profile is small enough and it is in the range of fairly large $z>3000$ where radiation is almost purely trapped due to small free path length both for scattering in lines of the main series and for Thomson scattering. So in the first case we may neglect an influence of dilution and Thomson scattering. In the second case this influence does not depend on the moment of source appearance as our calculations show \cite{dubr2011,dubr2015}. It should be stressed that frequency profiles of luminescent lines lines form at the moment of their creation (at some $z$) but their angular and spatial distributions form afterwards in the process of multiple scatterings on electrons on the way from the moment of creation to the moment of registration.

The considered problem can be "factorized" as follows. First, we obtained solution of space-independent problem: namely we calculated line profile of luminescent line for a source of continuum radiation homogeneously distributed in infinite space. Then we find the dilution factor and angular distribution of radiation due to multiple scattering on electrons for a source of finite size. Solution of space-independent problem gives an upper limit for luminescent line
intensity. Thus for the case of spherically symmetrical source an intensity of radiation in some subordinate line (transition $i\leftrightarrow j$) registered at present-day epoch ($z=0$) at the distance $r$ from a source center and at the angular distance $\theta$ from center direction may be written as
\begin{eqnarray}
\label{Inutr}
I(\nu,\theta,r)=I_{ij}(\nu)a(r)\chi(\theta,r),
\end{eqnarray}
where $\nu=\nu_{ij}/(1+z)$, $I_{ij}(\nu)$ is a solution of space-independent problem, $a(r)$ is a dilution factor, $\chi(\theta,r)$ is an angular distribution of radiation.

Space-independent distribution of radiation intensity is found from solution of nonstationary radiation transfer equation in a homogeneous expanding Universe which has the form
\begin{eqnarray}
\label{transfer}
H(z)\left[(1+z)\frac{\partial i(\nu,z)}{\partial z}+\nu
\frac{\partial i(\nu,z)}{\partial\nu}\right]=c\,k(\nu,z)[i(\nu,z)-s(\nu,z)].
\end{eqnarray}
Here $i(\nu,z)$ and $s(\nu,z)$ are dimensionless (measured in the mean occupation numbers of photon states) specific intensity and source function
respectively, $k(\nu,z)$ is a volume absorption coefficient, $H(z)$ is the Hubble factor. 

The formal solution of this equation - the solution for a given source function $s(\nu,z)$ in the righthand side was obtained in \cite{Dubrovich2004}. We consider radiation transfer in line (transition $i\leftrightarrow k$) assuming complete frequency redistribution (CFR) under scatterings. Then a source function does not depend on frequency ($s(\nu,z)\equiv s(z)$) and the formal solution in a narrow line approximation when a profile of absorption coefficient is replaced by delta function: $\varphi(\nu)=\delta(\nu-\nu_{ik})$, has the form \cite{Dubrovich2004}
\begin{eqnarray}
\label{izl}
i(\nu_{ik},z)=F(\nu_{ik}/(1+z))e^{-\tau(z)}+s(z)\left[1-e^{-\tau(z)},
\right].
\end{eqnarray}
where $\nu_{ik}$ is the laboratory frequency of atomic transition, $F(\nu/(1+z))$ is the initial (at $z=z_0$) radiation intensity and $ \tau(z) $ is the Sobolev optical length given by Eq. (\ref{tau}). 

Furthermore, we considered a problem of scattering of blackbody radiation (with temperature $ T + \Delta T $ for a fixed $ \Delta T/T $) on a spectral line in a medium with a temperature $ T = T_0 (1 + z) $. In this case for a given single-scattering albedo $ \lambda (z) $ the initial intensity of radiation and the source function are
\begin{eqnarray}
\label{account}
F(\nu_{ik}/(1+z))=b(\nu_{ik},T+\Delta T),
\end{eqnarray}
and
\begin{eqnarray}
\label{slamb}
s(z)=\lambda(z)\int_0^{\infty}\varphi(\nu)i(\nu,z)d\nu+[1-\lambda(z)]
b(\nu_{ik},T),
\end{eqnarray}
respectively. In Eq. (\ref{slamb}) the second term in the righthand side describes primary (thermal)
sources of radiation in a medium with the temperature $T=T_0(1+z)$ and
\begin{eqnarray}
\label{bnuT}
b(\nu,T)=\left(e^{h\nu/kT}-1\right)^{-1}.
\end{eqnarray}
In a narrow line approximation equation (\ref{slamb}) can be written as 
\begin{eqnarray}
\label{slam}
s(z)=\lambda(z)i(\nu_{ik},z)+[1-\lambda(z)]b(\nu_{ik},T),
\end{eqnarray}
Eliminating the source function $s(z)$ from Eqs. (\ref{slam}) and (\ref{izl}) and taking into account that $F(\nu_{ik}/(1+z))=b(\nu_{ik},T+\Delta T)$ in
Eq. (\ref{izl}) we obtain
\begin{eqnarray}
\label{DibP}
i(\nu_{ik},z)-b(\nu_{ik},T+\Delta T)=-[b(\nu_{ik},T+\Delta T)-b(\nu_{ik},T)]\,P(z),
\end{eqnarray}
where
\begin{eqnarray}
\label{Peq}
P(z)=\frac{[1-\lambda(z)][1-e^{-\tau(z)}]}{1-\lambda(z)+\lambda(z)\beta(z)}.
\end{eqnarray}
and $ \beta(z) = e^ {-\tau(z)} $. However, a more accurate consideration, which we do not present here because of its complexity, gives for $\beta$ the following
expression
\begin{eqnarray}
\label{beta}
\beta(z)=[1-e^{-\tau(z)}]/\tau(z).
\end{eqnarray}
We note that $\beta$ is the probability of photon escape from the process of scattering in an infinite homogeneously expanding Universe. The righthand side of Eq. (\ref{Peq}) consists of the multipliers with an obvious physical sence, namely they are: photon "death" probability per scattering, the probability of photon absorption in a medium and mean number of photon scatterings.

Furthermore we used equations obtained above to calculate the intensity of HeI subordinate lines arising as a result of continuum radiation absorption in the lines of the main series (transitions $1^1S-k^1P$, $k = 2,3,4$). In Section \ref{method} we describe how we calculated quantum yield $q_{ij}(z)$ in some HeI subordinate lines. Energy (in units of specific radiation intensity: erg/(cm$^2\cdot$ s$\cdot$ Hz$\cdot$ sterad)), absorbed in the line $1^1S-k^1P$ is given by the right hand side of Eq. (\ref{DibP}) multiplied by the factor $2h\nu_{1^1S,k^1P}^3/c^2$. Dividing this energy by $hc\nu_{1^1S,k^1P}$ we obtain photon concentration per unit frequency interval and unit solid angle. Multiplying this concentration by $4\pi$ and by the width of frequency distribution (which can be considered as Doppler one $\Delta\nu=\nu_{1^1S,k^1P}\,u/c$,  $ u=\sqrt{2k_{B}T/m_{\rm atom}} $) we obtain photon concentration in the corresponding line of the main series. Further we multiply this concentration by $q_{ij}(z)$ to find photon concentration in considered subordinate line. Finally, multiplying it by $h\nu_{ij}c$ and dividing by $4\pi$ and by $\Delta\nu=\nu_{ij}u/c$ we find the radiation intensity
in this line
\begin{eqnarray}
\label{Inm}
I_{ij}(\nu)=\left[B(\nu_{1^1S,k^1P},T+\Delta T)-B(\nu_{1^1S,k^1P},T)\right]
\\\nonumber
\times
P_{1^1S,k^1P}(z)q_{ij}(z),
\end{eqnarray}
\begin{eqnarray}
\nu=\nu_{ij}/(1+z)
\end{eqnarray}
Here $B(\nu,T)$ is the intensity of BBR radiation. We defined the profile of radiation
intensity in the line as $I/B\equiv I_{ij}(\nu)/B(\nu_{ij},T)$. Then according to Eq. (\ref{Inm}) we find
\begin{eqnarray}
\label{inm}
I/B=\left(\frac{\nu_{1^1S,k^1P}}{\nu_{ij}}\right)^3\frac{b(\nu_{1^1S,k^1P},T+\Delta T)-b(\nu_{1^1S,k^1P},T)}
{b(\nu_{ij},T)}
\\\nonumber
\times
P_{1^1S,k^1P}(z)\;q_{ij}(z).
\end{eqnarray}
Here $b(\nu_{ij},T)=b(\nu,T_0)$ according to Eq. (\ref{bnuT}) with $\nu=\nu_{ij}/(1+z)$ and $T=T_0$

The factor $P_{1^1S,k^1P}(z)$ in Eq. (\ref{inm}) depends on the single-scattering albedo $\lambda$ in $1^1S-k^1P$ line of the main series. Taking into account spontaneous and induced transitions in the field of blackbody radiation with the temperature $T$ we have
\begin{eqnarray}
\lambda_{1^1S,k^1P}=R_{k^1P,1^1S}/\left[\sum_{n=1}^{N}R_{k^1P,n^1S}+
\sum_{n=3}^{N}R_{k^1P,n^1D}+R_{k^1P,c}\right].
\end{eqnarray}
where $R_{k^1P,c}$ is the coefficient of photoionization from $ k^1P $ state and $ N $ is the number of considered states. 

\begin{figure}[ht]
\caption{Profiles of the $ 2^1P-2^1S $ line at $\Delta T / T = 0.01$ for the three pumping channels: $ 1^1S\rightarrow k^1P $ ($ k=2,3,4 $) and the total profile (sum curve).}
\centering
\includegraphics[scale=0.62]{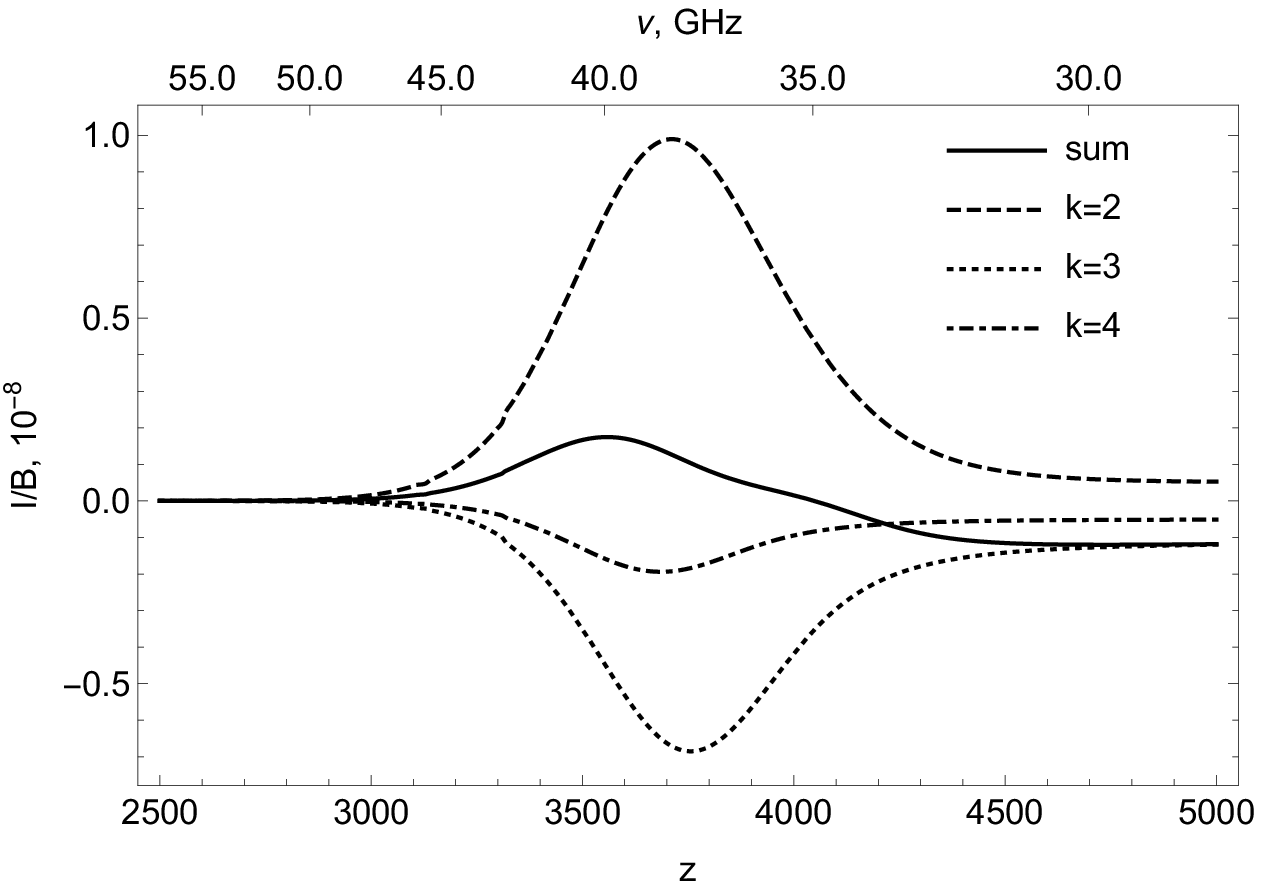}
\label{fig5}
\end{figure}

\begin{figure}[ht]
\caption{Profiles of the $ 2^1P-2^1S $ line at $\Delta T / T = 0.1$ for the three pumping channels: $ 1^1S\rightarrow k^1P $ ($ k=2,3,4 $) and the total profile (sum curve).}
\centering
\includegraphics[scale=0.62]{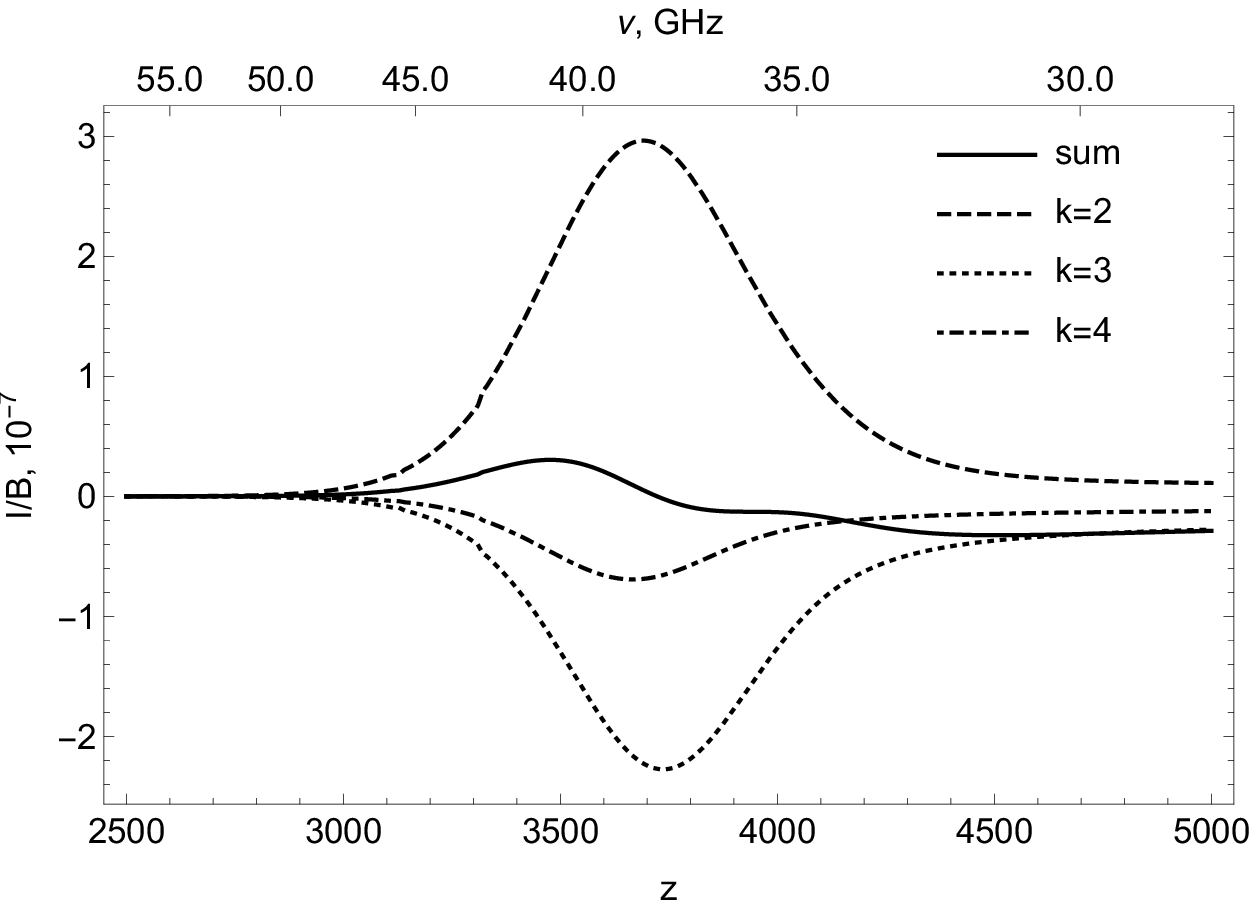}
\label{fig6}
\end{figure}

\begin{figure}[ht]
\caption{Profiles of the $ 2^1P-2^1S $ line at $\Delta T / T = 1$ for the three pumping channels: $ 1^1S\rightarrow k^1P $ ($ k=2,3,4 $) and the total profile (sum curve).}
\centering
\includegraphics[scale=0.62]{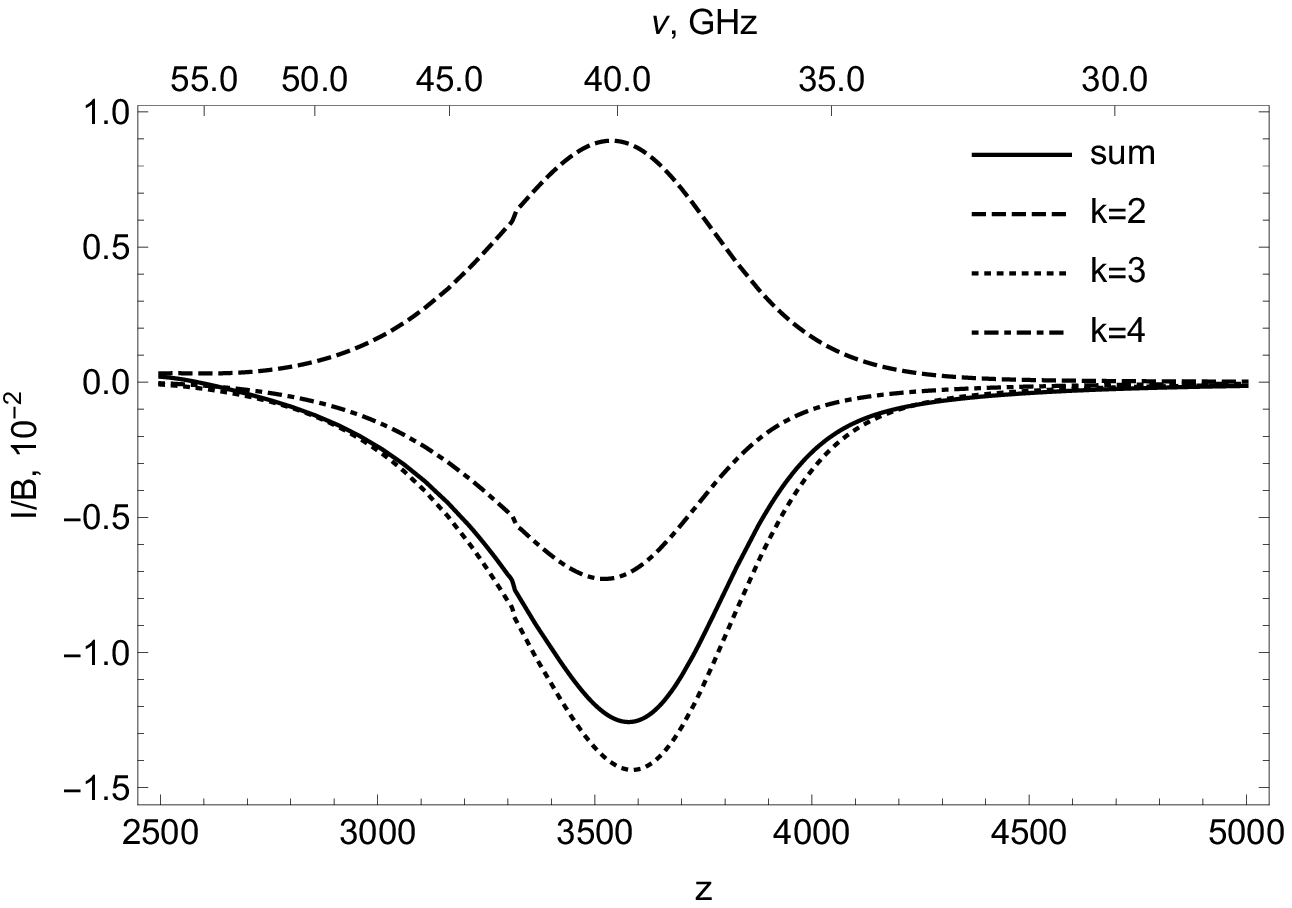}
\label{fig7}
\end{figure}

Calculations were performed using the cosmological parameters listed in Table \ref{param}. The profiles for the luminescent lines $1^1S-k^1P$ ($k = 2,3,4$) at different ratios $\Delta T/T$ are presented in Figs. \ref{fig5} - \ref{fig7}. 

We note the nonlinear dependence of profiles on the variation of the source temperature. It is interesting also to note transformation of a summary (with an account of all main pumping channels) profile from the emission profile to the absorption profile with the growth of $\Delta T/T$. This transformation points out the predominance of the pumping channels to the states $ 3^1P $ and $ 4^1P $ which make the considered line to be absorption one in contrast to the pumping channel $1^1S\rightarrow 2^1P$ (see Fig. \ref{fig1}). 

\begin{figure}[ht]
\caption{Spatial distributions of angle averaged radiation intensity at the present-day epoch for
the sources with different initial characteristic radiuses $r_{*}$ (in Mpc).}
\centering
\includegraphics[scale=0.62]{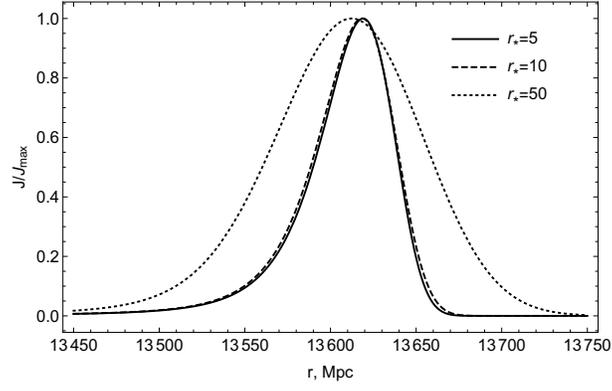}
\label{fig8}
\end{figure}

According to Figs. \ref{fig5}-\ref{fig7} the main part of the profile of the considered subordinate line is formed in the region $3000 <z <5000$ where as it was pointed above one may use Eq.  (\ref{Inutr}).
The problem of propagation of radiation burst with the account for multiple Thomson scattering on electrons in the expanding and recombining universe was solved earlier in \cite{dubr2016,dubr2011} where spatial and angular distributions of radiation intensity were calculated for spherically symmetrical sources with different initial sizes arising at different $z$. In Figs. \ref{fig8}-\ref{fig10} the spatial distributions of the angle averaged radiation intensity 
\begin{eqnarray}
J(r)=\frac{1}{2}\int_0^\pi I(r,\theta)\sin(\theta)d\theta
\end{eqnarray}
are presented. Figures \ref{fig9} and \ref{fig10} show the angular distributions $ \chi(\theta) = I(r,\theta) / I(r, 0) $ at the present day epoch ($z=0$) for the sources of different initial sizes (at the burst moment $z_0$), but for equal numbers of emitted photons. In the range $ z_0 = 3000 - 5000 $ these distributions are practically independent on $ z_0 $. From Fig. \ref{fig8} it is clear that spatial distributions of the radiation intensity for sources with sizes $ r_{*} = 5$  and $10$ Mpc have negligible difference. 
 Our calculations show that with the further diminishing of the size $r_{*}$ (but with the same number of photons initially emitted) spatial and angular distributions tend to a limit, which corresponds to the point source.

\section{Results and discussion}
\label{res}

In this paper we show that accounting for forbidden transitions significantly changes the luminescence intensity in para- and orthohelium and proved a very high efficiency on a wide set of spectral lines at large redshifts. The latter circumstance is very important for the reliable identification of the red shift of the object. In fact, this situation is close to what we have when observing distant quasars - the identification technology here can be similar. But the most important is the indication of the possibility of studying all these effects by ground-based radio astronomy.

\begin{figure}[ht]
\caption{Angular distributions of the radiation intensity $ \chi(\theta) $ at different distances $r$ (in Mpc) from the center of the source at the present day epoch. The characteristic radius of the source at the moment of the burst is $r_{*} = 50$ Mpc.}
\centering
\includegraphics[scale=0.62]{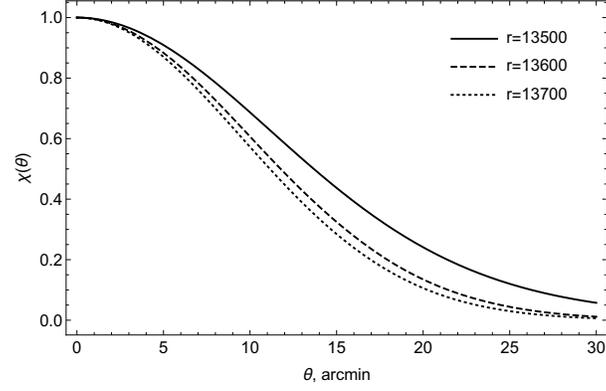}
\label{fig9}
\end{figure}

\begin{figure}[ht]
\caption{As in Fig. \ref{fig9} figure, but at $r_{*} = 10$ Mpc.}
\centering
\includegraphics[scale=0.62]{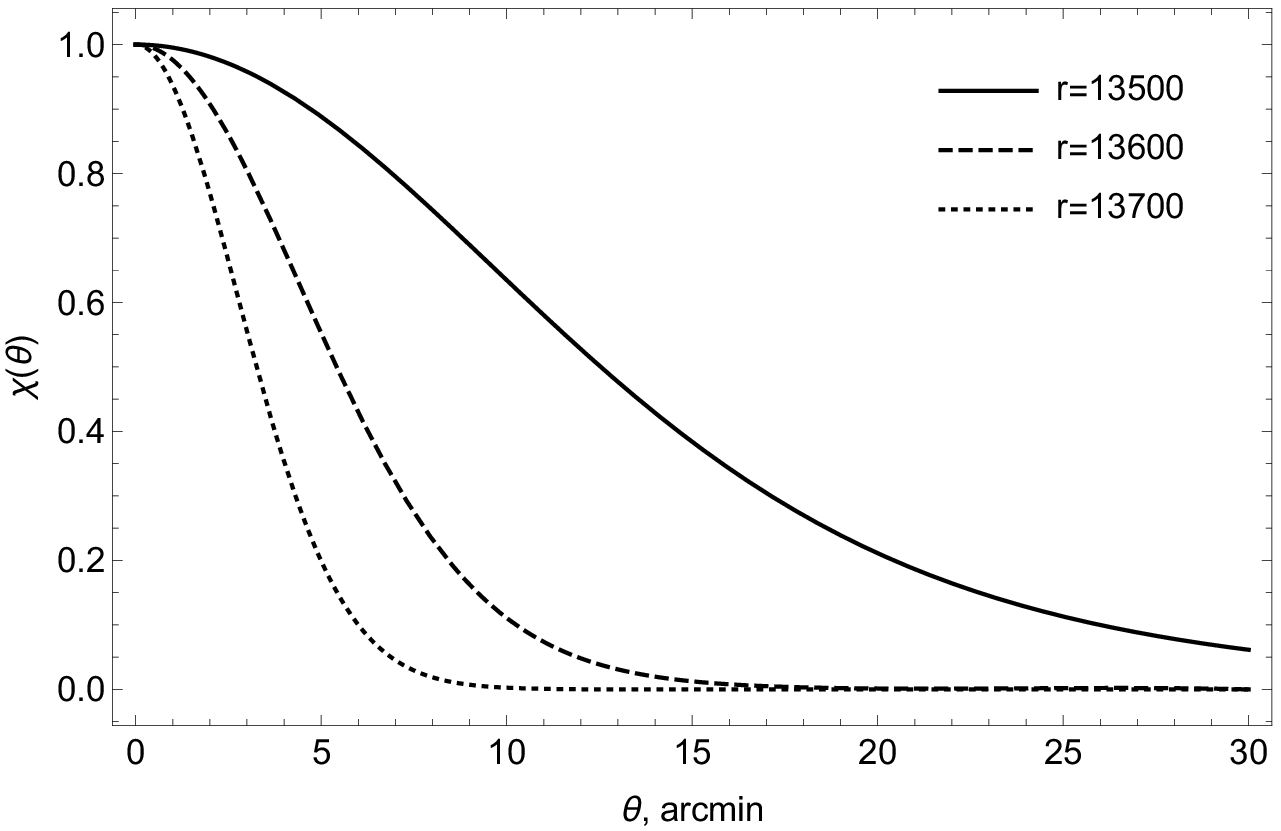}
\label{fig10}
\end{figure}

The spatial distribution obtained in this work describe photon distribution in number of scatterings. Photons were emitted at the moment $z_0$ and registered at the present day epoch $z=0$. If we neglect the width of the initial spatial distribution (for the sources with $r_{*}<10$ Mpc), all photons have gone the same distance and those who went by straight ways and therefore underwent less scatterings turned out to be ahead at the leading front of spatial distribution. These photons should have more narrow angular distributions than photons at the leading edge of spatial distribution as the Fig. \ref{fig10} shows. For the source of great initial size ($50$ Mpc) these differences are much smaller (see Fig. \ref{fig9}) and as a whole, angular diameter is much greater: $27$, $23$ and $22$ angular minutes on $r = 13500$, $13600$ and $13700$ Mpc. Dilution factor $a(r)$ appearing in Eq. (\ref{Inutr}) is equal $0.030$, $0.61$ and $0.069 $ respectively. For the source with $r_{*}=10$ Mpc (see Fig. \ref{fig10}) angular diameters on the same distances turns out to be smaller - $26$, $11$ and $7$ angular minutes and factor $a(r)$ is equal $1.2\times 10^{-4}$, $3.0\times 10^{-2}$ and $8.9\times 10^{-8 }$ respectively. For the source of $5$ Mpc size the same results are obtained. From what has been said above, it follows that for a sufficiently large source size with a sufficiently large $ \Delta T/T $, the luminescent lines can be very noticeable in the spectrum of blackbody background radiation.

\section*{Acknowledgments}
T. Z. acknowledges foundation for the advancement of theoretical physics "BASIS".

\end{document}